\documentclass[12pt,a4paper]{article}
\usepackage{times}
\usepackage{graphics}
\usepackage{graphicx}
\usepackage{float}
\restylefloat{figure}

\begin{document}
\large
\date{ }

\begin{center}
{\Huge  Experiments with ultracold neutrons -

\vskip 0.3cm

first 50 years.}

\vskip 0.7cm

{\Large Yu.N. Pokotilovski\footnote{e-mail:pokot@nf.jinr.ru}}

\vskip 0.7cm

              Joint Institute for Nuclear Research\\
              141980 Dubna, Moscow region, Russia\\

\vskip 0.7cm

{\bf Abstract\\}

\begin{minipage}{130mm}

\vskip 0.4cm

 This year is 50 anniversary of the first experimental observation of ultracold
neutrons (UCN).
 I present my reminiscences of the first experiment in Dubna,
first encounter with F.L. Shapiro -- pioneer of the UCN investigations, and the
early stage of the UCN experiments.
 Present status of investigations with UCN is shortly reviewed.

\end{minipage}
\end{center}

\vskip 0.3cm


\vskip 0.2cm


\vskip 0.6cm

\section{Prologue. My 1968 before UCN.}
 1968 was the last year of my postgraduate studentship in Moscow academic Institute.
The topic of my work was the experimental study of interactions of
high energy protons with big (quasi-infinite) uranium targets –- so called
electronuclear method of generation of neutrons and fuel isotopes.
The experimental part was performed at the JINR synchrocyclotron and
consisted in the measurement of production of neutrons and $^{239}$Pu
in big uranium targets of different depletion and different central
proton beam targets, bombarded by protons in the energy range 300-660
MeV.

 At that time this method was considered as a possible radical alternative to the
 breeder-reactor-plus-fuel-reprocessing industry and as a way to solve the future world
 energy problem due to involvement of all uranium to energy production without
 enrichment process.

 Later, when it became clear that situation with mineral sources of energy is not as
 sad as was predicted,  this theme was put aside.
 New life to this physics was done in nineties when previously (in sixties) discussed
 idea to use accelerators to burn out radioactive waste (nuclear waste transmutation)
 became of first priority in this method, and in a number of countries new numerous
 experiments are being performed to study transmutation.

 Experimental part of my work was mostly finished, two-three accelerator irradiations
of our uranium assembly and final calibration of fission detectors in
thermal neutron flux remained to be performed.
 Some experimental results from my thesis was published later\cite{elnucl}.

 The question was what to do after completing this work?
 I was not interested in my further participation in the field of elecronuclear breeding.
 On different reasons (scientific and personal) I was interested to continue my work
in Dubna and was recommended to contact Professor Fyodor L'vovich
Shapiro, vice-director of Laboratory of Neutron Physics (LNP) JINR.
 I do not remember was it the end of 1967 or beginning of 1968 when I knocked to the
door of his office.
 He immediately started to propose one theme after another.
 As I understood later he did not have at that time free people to test his ideas,
on the other hand he had second goal - to look if I am able to quickly enter to
new field.

 I must say that from the very beginning I was totally fascinated by extraordinary
personality of Professor Shapiro.
 My supervisor in my postgraduate studentshipness was talented, very active and
versatile scientist with interests from nuclear physics to chemistry
and biology.
 We usually met twice a year, ten minutes each meeting, in these ten minutes he solved
all practical problems, then - good bye.
 Discussions with Shapiro, usually deep at nights, were of quite opposite character: no
hurry, very quiet discussion with full attention to opinion of young interlocutor.

 The first proposal was the so called "molecular neutronoscopy"\cite{Gold} - electron
volt neutron scattering by molecules with the aim, in particular, to infer molecular force
constants from measured scattering cross sections.
 At energies essentially higher than the energy of chemical bond, nuclei that scatter
neutrons behave like quasifree targets, and it is possible to infer characteristics of
the interatomic potential from atomic momentum distribution.
 The  review of the  theory of this approach (impulse approximation) was published
in\cite{IvSa}.
 The inverse geometry time-of-flight spectrometry was assumed to be applied with
resonant detector -- neutron absorbing foil in the scattered beam
acting as a filter for the energy selection.
 This method was proposed by Shapiro et al. in 1961\cite{eV}.

 I quickly assembled the detector consisting of a hundred BF$_{3}$ counters, as a
resonance filter we used gold foil (E$_{res}$=4.91 eV).
 Laboratory of neutron physics had at that time 1 kW pulse reactor IBR.
Measurement with water sample showed that the method is hardly
sensitive, may be applied only to molecules with simplest structure,
and needs enormous statistics compared to what could be obtained from
IBR.

 Next idea proposed by Shapiro was to look more closely to the problem of limiting value
of proton polarization achieved in the crystallo-hydrate of
lanthan-magnesium nitrate La$_{2}$Mg$_{3}$(NO$_{3}$)$_{12}\cdot$
24H$_{2}$O.
 I was involved in collaboration with polish group working in inelastic thermal neutron
scattering (INS) on molecular crystals.
 The task was to estimate from measured torsional frequencies of the H$_{2}$O molecule
around the symmetry axes the value of the rotation barrier and from
the latter the magnitude of splitting of the ground state rotational
level in torsional potential.
 It was considered essential for understanding the limiting value of proton polarization
achieved in this substance.
 The splitting inferred from our measurements turned out not well determined: between
$2\times 10^{-17}$ and $5\times 10^{-10}$ meV\cite{pol}.

 In the end I myself performed short INS investigation on zeolitic water, one of the
first INS measurements of absorbed molecules\cite{zeo}.

 Then Shapiro proposed to think if it is possible to find Bose condensate in superfluid
helium by neutron scattering.
 The idea was based on recently published paper by Hohenberg and Platzman\cite{HoPl}.
 In this case simple estimates showed that existing LNF 1 kW reactor had too low power
to get any reasonable statistics in reasonable time.
 Thinking about this experiment I found in the book of Kerson Huang\cite{Kers}
reference to old paper\cite{Lamb} with the idea that helium Bose
condensation in momentum space should be accompanied by condensation
in configurational space - in the Earth gravity field Bose condensate
in liquid helium should be concentrated at the bottom of the sample.
 I still do not know if this effect was observed or investigated.

 After this Shapiro proposed me to think about the scheme of the experiment to observe
storage of ultracold neutrons.
 Half a year before I bought the book of Gurevich and Tarasov\cite{GurTar}, expecting
that low energy neutron physics will be my new field.
 I studied the chapters relating to slow neutron scattering but missed the chapter about
ultracold neutrons, considering them as pure exotic.
 After his introduction to this theme I said that it is the most exciting thing in low
energy neutron physics I ever heard, and that I am ready to put all my previous
deals aside and to start preparations to the experiment which seemed
to me not very complicated.
 After two-three discussions of possible schemes of the experiment and estimates of the
expected count rate he said: "I see you are burning with desire to start.
 You are a novice here, I decided to join other physicist - Strelkov - he works here
already about ten years and has good relations with our workshop".
 We quickly found common language and started preparations.


\section{1968 - first UCN observations.}
 As is known the first publication about possibility to store very low energy neutrons
in closed volumes was the paper of Ya.B. Zeldovich\cite{Zel}.
 In the first lines of his short paper Zeldovich writes:
"The idea of retaining slow neutrons has been mentioned many times,
but the corresponding experiments have not yet been performed, and
literature does not contain even rough estimates pertaining to this
problem".
 There was information from Bruno Pontecorvo that Fermi mentioned this
possibility long before but did not publish.
 Anyway Zeldovich was the first who published first estimates of the UCN reflectivity
and possible stored UCN densities.
 Later I have heard that after Zeldovich's publication possible experiments to observe
UCN were discussed in some Moscow neutron laboratories.
 But to my knowledge no attempt was made before our experiment.

 Important stimulus for performing the experiment on the UCN storage was the idea of
F.L. Shapiro\cite{ShaEDM} who proposed the UCN storage method to
search for the neutron electric dipole moment (EDM).
 Two advantages could be obtained in this method: decreasing the width of the magnetic
resonance due to increasing the neutron observation time in storage
volume compared to the cold neutron beam method, and second --
suppressing the so called ($v\times E$)-effect, imitating the neutron
EDM in case of not strong parallelism of electric and magnetic field.

 The experiment we performed in summer of 1968 was rather simple technically\cite{our-68}.
 I remind its main features.
 Our neutron source was the same 1 kW reactor IBR with frequency of pulses ~10 per
second.
 But as IBR was intended to be replaced next year by the new more powerful IBR-30
it was decided to try risky for the reactor mechanical stability regime but
necessary for our experiment: 6 kW  power with the  repetition rate 1 pulse
in 5 sec, increasing in this way the pulse power more than two orders
of magnitude.
 This regime of the neutron count rate measurement between well separated
pulses was crucial for this experiment.

 The scheme proposed by Shapiro was simple: the UCN guide, the UCN converter in
the beginning of the guide and some UCN detector at the end of the guide.
 The question was - how long should be the guide.
 We were afraid of strong UCN flux attenuation in their travel from the converter to the detector.
 Therefore we first tried short guide about 2 m long with detector placed in the reacor hall
close to the reactor.
 But it turned out that no available shielding of the detector could decrease the detector
background count rate to admissible level.
 Therefore finally we used long copper guide tube 10.5 m long with internal diameter 9.4 cm.
 As the UCN converter we used thin 1 mm polyethylene plate at the beginning of
the neutron guide, surrounded by massive paraffine moderator outside the guide.

  The UCN were counted by detector located at the end of the guide in intervals
between reactor pulses when all thermal and cold neutrons disappeared from the beam.
 Obviously, we could not use standard industrial $^{3}$He- or BF$_{3}$-filled neutron
detectors because their thick metal wall was practically a total UCN absorber.
 Thin scintillation detector ZnS(Ag) was prepared by sedimentation in water at the
surface of glass disc, after removing water and drying the scintillator was fixed by glue.
 Then the water solution of Li was deposited at the surface of scintillator, concentration
of Li isotopes was such that final substance at the surface of scintillator
LiOH$\cdot$H$_{2}$O had zero mean neutron scattering
length, and subsequently zero boundary energy for UCN.
 The neutron-sensitive layer optimized in efficiency for detecting the UCN was
several$\mu$m thick and had low efficiency to the thermal neutron background.
 It is interesting, that similar ZnS(Ag) scintillator but with Boron neutron sensitive layer
was used as an in-situ neutron detector in the very recent neutron lifetime measurement
in magnetic trap\cite{LA16,LA17}.

 Several arguments that we really observed the UCN storage:

 1. Time of flight information, demonstrating the constant count rate between the
pulses, e.g. long UCN dwell time in the neutron guide tube compared
to the interval between the reactor pulses.

 2. Thin 1.8 $\mu$m thick copper shutter closed alternatively one of two scintillation
detectors, reflecting UCN with energy below the boundary energy of
copper - 170 neV.

 3. We measured count rate with three different scintillator detectors:
 LiOH$\cdot$H$_{2}$O, LiF, and only ZnS(Ag) without neutron-sensitive layer.

 4. Experiments with addition of helium gas in the UCN guide to determine the UCN
dwell time in the guide.

 The count rate of the detector (copper shutter open mines shutter closed) was
 $\sim 7.5\times 10^{-3}$ s$^{-1}$.
 At the detector area 12 cm$^{2}$ and in the gas-kinetic assumption that the flux
density $\phi=\rho v/4$, where $\rho$ is the UCN density, and $v$=500
cm/s is the mean UCN velocity, we had $\rho=5\times 10^{-6}$
cm$^{-3}$ in vicinity of the detector (!).

 I was entrusted to present our experiment at the Laboratory seminar.
 The audience included some distinguished physicists from other laboratories.
 I remember the nagging questions of Bruno Pontecorvo about detector's efficiency and
background.

 Then it was decided to continue our UCN storage investigations at the more powerful
5 MW reactor IRT-M of Kurchatov Institute in Moscow in cooperation
with the group of Prof. L.V. Groshev.
 Shapiro asked me to make acquaintance with the place of future experiments and to
measure background in vicinity of the channel 3, where it was
proposed to install new UCN guide.
 Then we started preparations of new equipment for the experiments in Moscow.

 The same year Albert Steyerl constructed vertical neutron guide at the swimming pool
reactor in Garching.
 He provided it with chopper and measured in time of flight mode total neutron cross
section for aluminium and gold reaching the UCN velocity\cite{Ste1}.
 The UCN storage experiments he performed somewhat later\cite{Stebot}.
 In these storage experiments he observed interesting effect of significant albedo from
the walls of his graphite trap of neutrons with the energy higher than the
boundary energy of UCN  E$_{b}$ for graphite.
 He writes: "The results show that cooled graphite is very effective reflector material
for neutron energies up to 30 times of E$_{b}$".
 This observation was further investigated experimentally and theoretically and is
developed now to possible real technique to reflect and focus cold
neutrons (see below in the section on the UCN sources).


\section{The UCN storage problem.}
 The very first measurements of the UCN storage in summer of 1969 in Moscow started
to bring disappointment - the storage time in different UCN bottles: copper at
different chemical or electrochemical treatments, beryllium of
different form: foils or beryllium-covered stainless steel, pyrolytic
graphite, stainless steel, teflon, boron-free glass turned out to be
close\cite{myPLB71,my73,my76} and significantly lower than expected.

 The UCN loss factor $\eta$ inferred from the measured UCN storage times
was between $5\times 10^{-4}$ and $10^{-3}$.

 It was in strong contradiction with calculations for a pure perfectly smooth surface
based on known cross sections for these materials, predicting the UCN
storage losses for some of them one-two orders of magnitude lower.
 These calculations are based on simple theory $\eta=\sigma/(2\lambda b)$, where
$\lambda$ - the neutron wave length, $\sigma$ is the cross section of all
inelastic processes (neutron absorption and upscattering) at this wave length,
and $b$ is the coherent neutron scattering length.
 The experiments with heating and cooling the bottles (not very careful - I should say)
did not change the results.
 The observed independence of the UCN wall losses on temperature we explained by
"great compensation": heating removes absorbed water but increases inelastic
upscattering on the wall; cooling, on the contrary, decreases upscattering but
increases UCN absorption by  the hydrogen-containing molecules at the surface.

 Similar temperature independence of the UCN losses in copper and beryllium traps was
observed later by the group from the Leningrad Nuclear Physics
Institute in Gatchina\cite{Lob}.

 Two hypothesis were formulated very soon on the base of these observations: the surface
is covered with hydrogen containing film (most probably - water), or
there is some fundamental cause - anomalous propagation of sub-barrier neutron
wave into the wall violating our view of the UCN interaction with material potential barrier,
and more generally violating the quantum mechanics.
 Naturally two experiments were on the agenda:
a) to look - if the UCN upscattered by the wall nuclei or/and by
absorbed hydrogen and leaving the storage tube may be registered by
an external detector, and b) if the UCN propagate abnormally deep into
the wall.
 In the latter case two sorts of experiments are possible: propagation of
the UCN through thin foils and  the UCN activation experiment, with the activation
measured as a function of coordinate inside the wall irradiated by the UCN.

 UCN intensity was crucial for performing these experiments.
 In our first experiments in Moscow we first used aluminium at room temperature as
the UCN converter, then we tested some other UCN converters: water in
aluminium ampoule and zirconium hydride ZrH$_{x}$ with different
content of hydrogen\cite{my73}.
 Our UCN flux density at that time was, of course, much higher than in the first experiment
in Dubna - about 0.5 cm$^{-2}s^{-1}$, but was not sufficient to
perform these tests in full scale.

 The UCN transmission experiments through copper and teflon foils with thickness
$\sim 10 \mu$m did not show any propagation.
 I expected positive result of the UCN upscattering experiment: it is well known that
not outgasses surfaces contain absorbed gases.
 A number of different methods confirmed the presence of hydrogen, particularly in the
form of hydroxyl groups on metallic, metal-oxide and non-metallic surfaces.
 Water molecules are bound weakly (physisorbed) or strongly (chemisorbed), removal of
hydrogen from surface is not easy, occurs over wide range of
temperatures: from 100-200 C (physisorbed) to 1000 C (chemisorbed).

 The presence of hydrogen on the surface was demonstrated soon in direct detection of
hydrogen at a surface of copper, pyrolytic graphite and glass in
vacuum by the nuclear method\cite{LanGol} using resonant reaction
$^{15}$N$(p,^{12}$C)$^{4}$He plus 4.43 MeV gamma-ray.
 Different methods to reduce the hydrogen concentration on the surfaces are described
in\cite{LaMar}.

 I started preparations to the UCN surface activation measurements with the aim to find
abnormal UCN propagation to the depths larger 100-1000 \AA, having in mind that
10$ \mu$m may be too thick.
 What if abnormal propagation takes place only up to $\sim 1\mu$m?
 Electro-polished copper plates irradiated in the UCN flux were electrochemically etched
in different electrolytes, the thickness of the layer removed from the surface
being determined from the value of the electric charge passing through electrolyte cell,
then small samples were prepared after simple chemical procedure for the measurement
of $^{64}$Cu beta-activity.
 For the measurement of low beta-activity  the low background and high efficiency
miniature counter of beta-activity was constructed.
 The problem was to find the way to etch homogeneously (chemically or
electrochemically) the UCN irradiated copper plates layer by layer
with a resolution of some tens of  \AA.
 The first irradiations and measurements gave result for total activity in rough (20\%)
correspondence with the expected one for copper with obvious decrease
of activity with increasing depth of layer.
 Microscopic investigation of etched surfaces showed that homogeneity of etching was not
achieved.
 It was decided to use not differential but integral method: thin layers of copper were
deposited by thermal evaporation at the surface of float-glass
plates, after irradiation and preparation of the sample the activity
was measured as a function of total thickness of deposited copper.
 These experiments could be finished only much later after reconstruction of IRT-M reactor
to IR-8 and construction of new UCN guide with H$_{2}$O ice converter
with the UCN flux density about 3 cm$^{-2}$s$^{-1}$\cite{Act}.
 It was shown that there is no abnormal propagation of the UCN wave function at least at
the level $10^{-5}$.

 It should be mentioned that the idea of abnormal sub-barrier UCN propagation was
developed (mostly prompted by our observation of unexpectedly poor UCN storage)
by V.K. Ignatovich starting from\cite{Ignpac} and then in a number of
publications\cite{Ignpack}.
 The idea was based on non-spreading singular wave packets of the de Broglie
type\cite{deBrog}.

 After several years the scintillation UCN detector in our UCN storage experiments was
replaced by the proportional counter\cite{my73} (constructed by A. Strelkov) with better
pulse amplitude spectrum but having aluminium window reflecting and absorbing part
of the UCN spectrum.
 For better UCN detection efficiency it should be placed 50-80 cm lower than the UCN
storage bottle in order to accelerate the UCN by gravity before the
passage of 100 $\mu$m aluminium foil.

 The non-reflecting UCN scintillation detectors properly optimized\cite{mydet} were used
mostly in our UCN time-of-flight experiments.


 The first UCN upscattering experiment was performed by A.V. Strelkov et al. in
Dimitrovgrad\cite{StrHet} at much higher UCN flux.
 It was demonstrated that in not outgassed copper bottle at room temperature $\sim$75 \%
of the UCN lost in the bottle are upscattered.

 But what about outgassed UCN traps?
 What about other materials?
 Did our observation mean that all tested materials had the same hydrogen contaminations
leading to the same UCN upscattering?
 And what about temperature dependence?
 Important questions were still not answered.

 Estimates of the UCN upscattering by hydrogen in surface layers were made
in\cite{IgnSat,Blokh}.

 Temperature dependent behavior of the UCN losses in the temperature range 300-800 K
in the outgassed aluminium traps was found in\cite{MorAl}, but experimental losses were
still exceeding the theoretical ones more than an order of magnitude.
 Ion bombardment\cite{ion} and glow discharge\cite{Glow} cleaning of stainless steel,
aluminium and quartz bottles improved significantly the storage times.

 What strategy could be used to decrease the UCN losses in traps: to bake storage
chamber to high temperatures, to use different technologies of cleaning surface by
any accessible method like ion bombardment or glow discharge, or to try to cover
contaminated surface by hydrogen-free material, better at low temperature?
 The latter way was used in\cite{Kos91} and \cite{Alf761}, where D$_{2}$O and
CO$_{2}$ were condensed at the surface of storage copper chamber at 80 K.
 It was observed that losses practically did not change after this condensation.
 Better effect was obtained in\cite{Mor82}, but in all cases it was suspected
that condensed layers were porous and almost transparent for the UCN.


 New character the UCN loss anomaly acquired after the experiments in ILL with Be
trap at low temperature\cite{Anom1} and when the joint JINR-PNPI
group performed in Gatchina the UCN storage measurements in Be and
solid oxygen bottles at low temperatures\cite{Alf,Anom2}.

 The disagreement between experiment and calculations turned out to be the larger the
lower was the temperature of the walls.
 For example, the UCN wall reflection losses were exceeding by two orders of magnitude
the predicted ones from transmission cross sections of cold neutrons and from the
calculations based on the known phonon excitation spectrum of Be.
 At room temperature the wall losses in Be traps exceeded the calculated ones by an order
of magnitude.

 The table from\cite{anRev} summarizes these and other experimental data and
calculations.

\begin{center}
UCN loss coefficient $\eta_{storage}$ from UCN storage experiments
and $\eta_{theor,trans}$ from cold neutron transmission and dynamic
model calculations.
\end{center}
\begin{tabbing}
Substanqqqceqq\=qqqqqqqqqqqqqqq\=qqqqqqqqqqqqqqqqqqqqqqqqqqqqqqqqqq\=\kill
Substance     \> $\eta_{storage}$               \>$\eta_{theor,trans}$  \\
Be(6.5 K)     \> 3.2$\times 10^{-5}$\cite{Anom1}\>3$\times 10^{-7}$
(Debye model calc.)                                                    \\
Be(300 K)     \>  4$\times 10^{-5}$\cite{Anom2}  \>5$\times 10^{-6}$
(cold neutron cross sections\cite{Mug})                                           \\
Be(10 K)      \> 3.0$\times 10^{-5}$\cite{Anom2}\>3$\times 10^{-7}$
(Debye model calc.)                                                       \\
O$_{2}$ (10 K)\> 6$\times 10^{-6}$\cite{Anom2}  \>6$\times 10^{-7}$
(magnon spectrum calc.\cite{oxy,oxyliu})                                 \\
C (100 K)     \> 5$\times 10^{-5}$\cite{Arz1}   \>2$\times 10^{-6}$
(cold neutron cross sections\cite{Mug})                                             \\
D$_{2}$O(80 K)\>9.4$\times 10^{-6}$\cite{Arz1}  \>$\leq 2\times
10^{6}$
(cold neutron cross sections\cite{Gissl,Uts}                              \\
D$_{2}$O(90 K)\>$\sim 6\times 10^{-5}$\cite{Benew} \>$\leq 2\times
10^{6}$
(cold neutron cross sections\cite{Gissl,Uts}                               \\
D$_{2}$O(7 K) \>$\sim 6\times 10^{-5}$\cite{Benew}   \>$\leq 2\times
10^{6}$ (cold neutron cross sections\cite{Gissl,Uts}
\end{tabbing}
 No essentially new results appeared for these materials from that time up to now.

 It was natural to continue studying the UCN upscattering to thermal and cold neutron
energy range in more details for different materials and different temperatures.
 After numerous difficult experiments the summary picture looks rather controversial,
 I can only give here some references on the experimental publications\cite{upsca}.

 Theoretical investigations of the UCN interactions with matter are contained
in\cite{Blokh,Barab}.


 New important step in the UCN storage was made due to the proposal\cite{Bates}
and application of liquid hydrogen-free fluoropolymers --
perfluoropolyether (PFPE) oils\cite{Solway} -- for covering the walls of traps.
 The UCN loss coefficient at room temperature in liquid oil (Fomblin)
traps approached  $\eta\sim 2\times 10^{-5}$\cite{Rich}.

 The UCN traps with liquid walls were used for the neutron lifetime
measurements\cite{MAM00,MAM10}.
 It was shown that in liquid phase, the quasielastic UCN scattering by viscoelastic
surface waves\cite{surf1,Lamor} is important and possibly the main
cause of the UCN losses in liquid traps.

 Lower UCN losses were expected with the low-temperature modification of
fluoro-polymers\cite{LTF1}, which have much lower pour point ($\sim$ -100 C) in
comparison to Fomblin.
 The measured\cite{LTF1} cold neutron cross sections promised rather small low-temperature
(-100$^{o}$\,C ) UCN loss coefficient - about $2\times 10^{-6}$.
 It was confirmed later in the neutron lifetime measurement\cite{Ser05} where this
fluoropolymer was used in solid state at lower wall temperatures -
down to -160 $^{o}$ C.

 As the calculated neutron capture loss coefficient for these substances is almost
order of magnitude lower ($\sim 3\times 10^{-7}$) it seems that the main cause of losses is
still the UCN upscattering.
 It is supported by the direct measurements of density of states and the calculations on
this base of the UCN loss coefficient\cite{DOS}.

 It became clear from all previous experience that for better UCN storage the surface
should consist of nuclei with low neutron capture (the choice is not
wide: D, Be, C, O, F), to be at low temperature and ideally smooth.
 The latter is hardly achieved with frozen C, D$_{2}$O, CO$_{2}$, Be, O$_{2}$ etc,
and the only hope in my view consists in further developing technology of
preparation of perfect surfaces of fluoropolymers.

 The final goal in the UCN storage problem - to suppress upscattering losses completely
and to reach the value typical for fluoropolymers $\sim 3\times 10^{-7}$ determined
by nuclear capture.
 Is this goal achievable?
 New experiments are necessary.


 Suspicions about the possibility of small energy changes in UCN at wall reflections in
the traps were voiced long ago but without indicating any physical
mechanism (see, for example,\cite{Fr}).
 The effect of possible wall sound vibrations was estimated in\cite{Ger}.
 The low frequency part of the phonon spectrum of solids and the existence of low
frequency vibrating clusters in disordered solids were considered
in\cite{Ign}.

 The experimental search for small energy transfer at the UCN interaction with the walls
of storage volume was the topic from the early stage of the UCN
storage experiments\cite{MorAl}.
 Later searches are described in\cite{Stesmall}.

 In a number of more recent experiments it was shown that there are small
($\sim 10^{-7}$ eV) UCN energy transfers at the UCN reflection from
the walls of storage volumes\cite{small1}-\cite{small9}.
 The reported probability of this effect per reflection was from
$\sim 10^{-5}$ in the first publications for metal surfaces and for
the surface of liquid fluorinated oil (Fomblin), to $\sim 10^{-7}$ in
more recent publications.
 The UCN down-scattering with lower probability has been reported in\cite{Morc,Mor}.

 But in the experiments\cite{Ser,Ser1} it was found that the probability
of "small heating" at UCN reflection from solid walls is $\sim
10^{-8}$ or lower, the probability of the effect at the reflection
from viscous liquid fluorinated oil being of order of $10^{-5}$ -- in
good agreement with the previous observations and calculations
of\cite{surf1}.
 The experimental setups and the methods of measurements in\cite{small1}-\cite{small9}
and\cite{Ser,Ser1} were close, the main contradiction concerned the
efficiency of detection of upscattered neutrons.
 In both setups the upscattered neutrons experienced many collisions
with the walls of the storage volume before they could reach the detector.
 This sets the upper energy boundary for the detected neutrons and
affects the detection efficiency of upscattered neutrons.

 The search for the UCN upscattering from beryllium surface to $\mu$eV energy range
by the method of indium foil activation method was performed in\cite{Inact}.

 The method to detect and measure spectra of weakly upscattered UCN, different
from\cite{small1}-\cite{small9} was tested in\cite{mysmal}.

 The low-energy upscattering from liquid surface was described as quasi-elastic neutron
scattering by viscoelastic surface waves \cite{surf1,Lamor}.
 As for the solid reflecting surface the nature of the effect is not quite clear.
 The probability of phonon upscattering with very small neutron
energy changes is many orders of magnitude lower.

 Two mechanisms were proposed to explain this extraordinary large scattering
probability to small neutron momentum space volume at reflection from
solid surface: diffusive motion of absorbed and dissolved
hydrogen\cite{hydr}, and the motion of nano-particles at the
surface\cite{Ignclu,nesvclu}.


\section{UCN production and sources.}

 As was said in the first experiments in Kurchatov Institute\cite{my73} we tested Al,
water in aluminium ampoule and zirconium hydride ZrH$_{1.9}$ as the UCN converters,
and then additionally magnesium in Alma-Ata\cite{A-Ata} with the UCN flux
density in the latter case about 1 cm$^{-2}$s$^{-1}$ .
 The UCN flux density of $\sim 20$ cm$^{-2}$s$^{-1}$ was obtained in\cite{SM2}
at the Dimitrovgrad high power 100 MW reactor SM-2 (thermal neutron flux density at the
converter $\sim (2-4)\times 10^{14}$ cm$^{-2}$s$^{-1}$) and water cooled
zirconium hydride converter.
 Order of magnitude increase in the UCN production ($\sim 200$ cm$^{-2}$s$^{-1}$) was
achieved  in LNPI (Gatchina) at the vertical guide when Be converter
was cooled down 30 K\cite{GatBe}.

 New era of the UCN production started after putting in operation in Gatchina (LNPI) in
1980 of liquid hydrogen source\cite{Ser80} and in 1986 of Steyerl
turbine in ILL\cite{Stesource}.
 This latter UCN source is the world UCN source used by a multitude of the UCN groups
during more than 30 years.

 Approximately the same time the systematic investigations were started of solid
deuterium and liquid helium as the most effective materials for the UCN production.

{\it Solid deuterium.}

 The first estimate of the UCN production and upscattering cross sections in
solid deuterium was made in\cite{D2-Gol}.
 The phonon frequency spectrum of solid deuterium was calculated in\cite{Gol-D} using
the dispersion relations measured in \cite{disp} (see also\cite{Schm}).

 The first test of the UCN production in solid deuterium has been performed by
the PNPI group\cite{Ser80}, which then has been studying experimentally the UCN
production in deuterium during many years\cite{Ser-86}-\cite{Ser-01}.

 Later calculations in the incoherent approximation\cite{incoh} of the UCN
scattering on phonons in solid deuterium confirmed the results of
\cite{Gol-D}, this publication also contains the calculation of the
UCN spin-flip upscattering on para-deuterium with para-ortho
transitions of deuterium molecules.
 Validity of incoherent model for calculations of neutron interaction in solid deuterium
and the UCN generation was studied in the works\cite{Gra,Lav,myPSI}.
 The incoherent approximation and the density of states of\cite{Gol-D} was
used in the calculations of the UCN production at the TRIGA reactor
in Mainz\cite{my-calc}.
 Additional detailed measurements of dynamic properties of solid deuterium were
performed at the FRM-2\cite{Gut}.

 Transmission cross sections of liquid deuterium were measured in\cite{PSIliq,Doeg},
and of solid deuterium in\cite{PSIsol}.

 Systematic investigation of the UCN production in gas, liquid and solid phases
of deuterium, oxygen and deutero-methane was performed in
PSI\cite{myPSI}.

 Homogeneity of solid deuterium crystals is very essential for effective UCN
extraction because inhomogeneities decrease the UCN free path length and
respectively efficiency of extraction.
 Preparation procedure and temperature treatment of solid deuterium crystals were the
subject of\cite{Ser-01,PSI-light,sD-cross}.
 Monte Carlo simulations of the UCN transport and extraction from inhomogeneous solid
deuterium crystals is contained in\cite{mytrans}.

 It was proposed that because of comparatively short neutron lifetime in solid deuterium
(140 ms in pure ortho-deuterium at T=0 K and decreasing because of thermal
upscattering on phonons and admixture of para-fraction and absorption by
admixture of hydrogen) solid deuterium can find better application in a pulsed
mode of the UCN production\cite{prod}.
 The UCN transport simulation in solid deuterium crystal with fast changing temperature
relevant to pulsed UCN production was performed in\cite{gran}.
 This situation is realized when pulse width of the neutron source is
comparable with the time of the UCN transport in converter.
 TRIGA reactor with its pulse width of $\sim$30 ms is a typical example.

 General idea of using pulsed neutron sources for the UCN
production consists in pulse shutter in close vicinity to the UCN
source: closing the UCN source from the neutron guide between pulses
prevents absorption of the UCN in the source.
 This is hardly realizable technically at the pulse channel reactors.
 But for the neutron sources with well separated pulses (e.g. TRIGA type pulse reactors)
the shutter may be placed at the end of the guide, at the entrance to the UCN storage
volume\cite{prod}.
 UCN produced in the moderator-converter during the pulse spread over the mirror
neutron guide.
 The UCN reaching the storage volume spend in it some time before leaving back to the
neutron guide, by this time all the tail of the UCN reach the trap and may be captured
from the guide in the trap.
 The fast shutter located near the entrance window of this volume should be closed
at the proper moment after the pulse.
 This was tested in\cite{Mainz}.

 Now after years of preliminary investigations the solid deuterium UCN sources are in
use in Mainz\cite{Mainz}, LANL\cite{LA} and according to
proposal\cite{Ser-PSI} is realized in PSI\cite{PSI}.
 The UCN sources with solid deuterium converter are in stage of construction at
the FRM-2\cite{FRM} and in North Caroline University\cite{Puls}.

{\it Liquid helium.}

 The first proposal and calculation of the UCN production in liquid helium was published
by Golub and Pendlebury\cite{He1}.
 In very low temperature (below 0.5 K) liquid helium the UCN upscattering due to
one-phonon absorption is suppressed.
 Therefore in spite of small cold neutron downscattering cross section and ,respectively,
low UCN production efficiency high density UCN gas can be accumulated in result of long
irradiation of a helium bath by high intensity cold neutron beam.
 Numerous calculations of neutron-liquid helium interaction cross
sections\cite{He2}-\cite{Abe01}, liquid helium heating by the
neutron and gamma radiation\cite{Yosh87}, the UCN production
probability in liquid helium\cite{He4,Abe01,Abe03,HeKor} were
conducted.
 The UCN production measurements were performed in \cite{He3}-\cite{Jap12}.

 The liquid helium UCN sources are under construction in Japan\cite{HeJap},
in ILL\cite{HeNesv} and is planned in PNPI\cite{HeSer}.

 New intense UCN source based on cold neutron moderation and storage in
liquid helium at a dedicated beam line at ILL is constructed by Zimmer et
al.\cite{Zimm07}-\cite{Zimm16}.

 In the paper\cite{Laminh} authors discussed a possibility to trap cold neutrons to
increase the UCN production in liquid helium bath.
 The idea is based on Steyerl observation of significant albedo of superbarrier neutrons
from inhomogeneous medium\cite{Stebot}.
 Among other materials for reflectors they considered BeO, for which they expected
strongly enlarged scattering cross section due to its grainy structure.
 This possibility was further experimentally investigated in\cite{Morinh} for a number
of inhomogeneous mixtures.
 Neutron reflectivity of inhomogeneous structures was calculated in\cite{Art}.
 Detailed investigations of reflectivity of nano-diamond powders was performed by
Nesvizhevsky et al.\cite{Nesinhom}.


\section{Neutron decay.}

{\it Neutron lifetime.}

 As is well known the value of the neutron lifetime is important for obtaining the
Standard Model parameter - the CKM matrix element $V_{ud}$ and in
astrophysics - for understanding the Big Bang Nucleosynthesis
\cite{Dubb,Witf}.

 There are two different experimental methods of measurement the neutron lifetime.
 One approach - the beam method - consists in measuring number of neutrons in a
well-defined volume of a neutron beam and simultaneously the count
rate of the neutron decay in this volume through detection of
products of the neutron decay - electrons or protons.
 Two absolute measurements are performed in this method: neutron density in the
beam and number of counted protons.

 The second method consists in the UCN storage in material or magnetic\cite{Vlad}
traps and measuring the number of remaining neutrons as a function of
time.

 The first attempt to apply the UCN storage method for measurement the neutron lifetime
was made by V.I. Morozov et al.\cite{Mor1,Mor2}.
 They used aluminium vertical cylinder as a storage bottle at room temperature\cite{Mor1}
and at liquid nitrogen temperature\cite{Mor2} with subsequent results
for the neutron lifetime 875$\pm$95 s\cite{Mor1}, 903$\pm$13
s\cite{Mor2}, and 900$\pm$11 s\cite{Mor3}, respectively.
 The main problems in these pioneer experiments were low UCN intensity at their UCN
source in Dimitrovgrad, narrow energy range of stored neutrons, and relatively large
UCN losses in collisions with walls of the storage volumes.

 After construction of intense UCN sources in Gatchina (liquid hydrogen)\cite{Ser80}
and in ILL (famous Steyerl turbine)\cite{Stesource} and decreasing UCN losses in wall
collisions in traps, the precision of the UCN method was raised to $\sim$3 s.
 Joint group of JINR (Dubna) and PNPI (Gatchina) stored UCN in the low temperature (15 K)
beryllium and solid oxygen gravitational trap\cite{Gat0,Gat1,Gat2}.
 The ILL group used for covering the walls of storage volume the commercial
perfluorinated oil (FOMBLIN)\cite{MAM1}.

 Last 20 years of the neutron lifetime measurements brought new problem.
 There is disagreement between the beam and the UCN storage methods of
determination of the neutron lifetime in the most accurate measurements of
the last two decades.

 A summary of recent neutron lifetime measurements: two
beam experiments\cite{Byrne,NIST05} (correction of the latter in
\cite{NIST13}), five UCN storage experiments in material
traps\cite{MAM00,MAM10,Ser05,Ser17} and\cite{Arz00} with the
subsequent corrections of the latter\cite{Arz12,Arz15}, and the first
successful UCN magnetic storage neutron lifetime experiment
\cite{Ezh09,LA16,LA17} gives:
the average of two neutron beam experiments (after correction\cite{NIST13}) gives
$\tau_{n}=888.0\pm 2.1 s$, the same for six latest resuts obtained by the  UCN storage
method (after two corrections\cite{Arz12,Arz15}) gives $\tau_{n}=879.17\pm 0.4
s$.
 The disagreement between two methods is more than four standard deviations.
 Comments concerning experiments\cite{MAM00,Ser05} are in\cite{Ste-rough,Ste-liq}.
 The most probable cause of this difference are systematic errors in either of
these two methods.

 According to\cite{NIST05} in the beam experiment the main uncertainty comes
from determination of the cold neutron flux.
 But if some small part of protons avoids detection - this causes
overestimation of the neutron lifetime.
 In the UCN method possible poorly controlled neutron losses in traps lead to
lower value of the measured neutron lifetime.

 Much more exotic scenarios were discussed as a cause of neutron
disappearance from the UCN traps: neutron - mirror neutron
oscillations\cite{Berezh}, or neutron propagation to brane world
\cite{bran} (neutron- shining-through-a-wall experiments).

 Nowadays the magnetic UCN storage for the precision neutron lifetime measurement
is more popular than UCN storage in material traps.
 Apart from current experiments\cite{Ezh09} and\cite{LA16} several new experiments
are in preparation: the experiment of Ezhov group with larger magnetic bottle\cite{Ezhnew},
the experiment with the UCN in magnetic bottle filled with liquid helium\cite{He-mag},
the PENELOPE experiment at the new solid deuterium UCN source at FRM-2\cite{PENEL},
the experiment of Zimmer group in ILL\cite{Zim-mag}, and the experiment in Mainz, using
spectrometer aSPECT\cite{aSPE}.

{\it Neutron decay correlations.}

 The first experiment to measure neutron beta decay asymmetry parameter $A_{0}$
(neutron spin -- electron momentum) was measured by the collaboration
from USA universities (UCNA Collaboration).
 In result of many years efforts the precision was considerably improved from
$A_{0}=-0.1138(46)_{stat}(21)_{syst}$\cite{UCNA09} to \\
$A_{0}=-0.12054(44)_{stat}(68)_{syst}$\cite{UCNA18}.


\section{Neutron EDM.}
 Search for the neutron EDM seems the most fundamental and complicated
experiment in the low energy fundamental physics.
 The first who realized the method proposed by F.L. Shapiro\cite{ShaEDM} was Gatchina
group of V.M. Lobashev\cite{EDM1} with the result d$_{n}<1.6\times
10^{-24}$ e$\cdot$cm, two times more strict limit than the latest
cold beam experiment\cite{Dress}.
 Next 35 years may be considered as kind of competition between Gatchina group
led by V.M Lobashov, and later by A.P. Serebrov  and the
international group, lead by J.M. Pendlebury with subsequent more and
more better constraints on the value of the neutron EDM:
d$_{n}<6\times 10^{-25}$ e$\cdot$cm\cite{EDM2},
d$_{n}<6\times 10^{-25}$ e$\cdot$cm\cite{EDM3},
d$_{n}=(0.3\pm 4.8)\times 10^{-25}$ e$\cdot$cm\cite{EDM4},
d$_{n}=(3\pm 5)\times 10^{-25}$ e$\cdot$cm\cite{EDM5},
d$_{n}<9.7\times 10^{-26}$ e$\cdot$cm\cite{EDM6},
d$_{n}<6.3\times 10^{-26}$ e$\cdot$cm\cite{EDM7},
d$_{n}<2.9\times 10^{-26}$ e$\cdot$cm\cite{EDM8},
d$_{n}<3\times 10^{-26}$ e$\cdot$cm\cite{EDM9},
d$_{n}<5.5\times 10^{-26}$ e$\cdot$cm\cite{EDM10},
 - two orders of increase in sensitivity compared to the neutron EDM beam experiments.

 The UCN EDM experiment became very popular.
 In recent report Klaus Kirch gives the list of new projects in different
countries\cite{Kirch} (apart from operational PNPI (Russia)-ILL
(France) and PSI (Switzerland)): RCNP (Japan) - TRIUMF (Canada),
FRM-2 (Germany), SNS (USA), PNPI (Russia), LANL (USA), WWR-M
(Russia), and three in more far future: PIK (Russia), J-PARK (Japan)
and ESS (Sweden).

\section{Other experiments with UCN.}

{\it Neutron optics with UCN.}

 First neutron optical experiments with the UCN were performed by Steyerl et al.
 They constructed the UCN gravity diffractometer\cite{Stedifr} and demonstrated
neutron quasistationary levels in two-hump potential barrier
structure (interferential filter)\cite{Stefilt}.
 Energy splitting due to tunnelling in coupled resonators was observed in\cite{Stefilt1}.
 The UCN transmission bands in multilayer structures were demonstrated in\cite{Steband}.

 Image formation with neutrons using the UCN focusing by reflecting concave
Fresnel zone mirrors in vertical geometry was obtained with reported
magnification about 6\cite{Stemag6}, up to 50\cite{Stemag50}, and
then up to 240\cite{Stemag240}.

 Horizontal UCN microscope was developed in\cite{Frankmicr}.

 It was predicted in\cite{metre} that in analogy with electromagnetic optics the so
called neutron "metallic reflection" from strongly neutron absorbing
matter exists for neutrons.
 The effect is essential when the path length of neutron in a medium is comparable or
less than its wavelength.
  The effect was observed in\cite{metref} for isotope $^{157}$Gd for which the real
part of the neutron scattering length is much less than the imaginary one.
 Neutron reflection in this case is due to the latter.

 Frank et al. performed a number of neutron optics experiments specific for the UCN
energy range: UCN diffraction on moving gratings\cite{Frankmov}, test
of $1/v$ law of the neutron absorption cross section in strongly
absorbing medium\cite{Frankabs} (see also experiments\cite{Rauchabs}
and comment\cite{abscom}), effect of accelerating matter on the UCN
energy change\cite{Frankacc}, test of the dispersion law for the
neutron waves\cite{Frankdisp}.

{\it Neutron $\rightarrow$mirror neutron oscillations.}

 The idea of possible existence of mirror world - where mirror particles were proposed
to restore the symmetry between left and right - was first mentioned
by Lee and Yang in their famous paper\cite{LeeYang}.
 See also the development of this idea in\cite{Kobz} and in a number of papers of Foot
with a review in\cite{Foot}.
 The history of the idea and its development is contained in a review of Okun\cite{Okun}.

  Possibility of the neutron $\rightarrow$mirror neutron oscillations was conjectured
by Berezhiani et al. in a number of recent publications\cite{Berezh}.
 Possible dedicated neutron beam and the UCN storage experiments and their sensitivity
were considered in\cite{Pokmir}.
 The first experiments for the search of  the neutron $\rightarrow$mirror neutron
oscillations with the UCN storage are described in\cite{Sermir} and\cite{PSImir}.
 The limits obtained in these experiments for the oscillation time was about 450 s.

{\it Neutron electric charge.}

 Possibility of nonzero neutron electric charge was discussed from different points of
view\cite{ncharge}.
 The best cold neutron beam experiment\cite{Baum} limited the neutron charge at the level
$q_{n}<1.8\times 10^{-21} q_{e}$ (95\% c.l.).

 Kashukeev et al. devised the UCN mirror camera with potentially better sensitivity
of the neutron trajectory to external fields\cite{Kash}.
 They tested this neutron optical instrument in applied electric field in the experiment
at WWR reactor in Gatchina\cite{Borcharge}.

 Further development of this interesting approach was described in\cite{Plo}
but desired sensitivity is still not reached and a full-scale experiment is still not performed.

{\it Neutron quantum levels in the Earth gravitational field.}

 First observation of  the neutron quantum levels in the Earth gravitational field is
described in\cite{1grav} (more recent review of similar experiments in\cite{gravrev}) .
 The principle is based on known quantum mechanical problem of quantum levels
of massive particle in the Earth gravitational field\cite{gravlev}.

 Resonance transitions between the first and higher gravitational levels by means of
mechanical oscillations was demonstrated in\cite{gravres}.
 More refined approach to perform resonance transitions is described in\cite{reschirp}.

 Experiments on resonance transmission between neutron gravitational levels
are sensitive to the search for new hypothetical short range
interactions\cite{gravconstr}.

{\it Search for Lorentz invariance violation.}

 High sensitivity UCN EDM spectrometer was used for the search for the directional
dependence of the neutron magnetic resonance frequency\cite{LorUCN}.
 The latter could be a consequence of existence of cosmic electric dipole tensor (Lorentz
invariance violation).

\section{Conclusion}
 Great number of methodical developments relating to the UCN physics were not
mentioned in this short review: high quality UCN guides, detectors, including
the high resolution coordinate ones, the questions of the UCN depolarization in
traps, quantum mechanical effects in the UCN behavior in the UCN EDM
installations etc.

 Several reviews were published in different times\cite{Revs} and two books\cite{Books}
covering the state-of-art of the investigations in this field contemporary to
the day of writing.

 The main problem for widening the UCN applications,
for example to condensed matter research (but the UCN storage problem is
actually the problem of the UCN interaction with matter) is the low UCN fluxes -
maximum $\sim 10^{4}$ cm$^{-2}$s$^{-1}$.
 New, non-standard ideas are needed to increase essentially the UCN fluxes.

\end{document}